\documentclass [noshowpacs,byrevtex,amssymb,superscriptaddress,aps,prl,floatfix,twocolumn]{revtex4}

\usepackage{amsmath}
\usepackage{amssymb}

\usepackage{mathrsfs}
\usepackage{eucal}
\usepackage{mathbbol}
\usepackage{epsfig} 
\usepackage{graphicx}
\usepackage{color}
\usepackage{epstopdf}

\begin{document}

\title{Complete dynamics of $\text{H}_2^+$ in strong laser fields}
\author{J. Handt}
\author{S. M. Krause}
\author{J. M. Rost}
\affiliation{Max Planck Institute for the Physics of Complex Systems, 
N\"othnitzer Stra{\ss}e 38, D-01187 Dresden, Germany}
\author{M. Fischer}\email[]{E-mail: Michael.Fischer@tu-dresden.de}
\author{F. Grossmann}
\author{R. Schmidt}
\affiliation{
 Institut f\"{u}r Theoretische Physik, Technische Universit\"{a}t 
Dresden, D-01062 Dresden, Germany}
\date{\today}

\begin{abstract}
Based on a combined quantum-classical treatment, a complete study of 
the strong field dynamics of H$_2^+$, i.e. including all nuclear and 
electronic DOF as well as dissociation and ionization, is presented.
We find that the ro-vibrational nuclear dynamics enhances dissociation 
and, at the same time, suppresses ionization, confirming experimental 
observations by I. Ben-Itzhak {\it et al.} [Phys. Rev. Lett. {\bf 95}, 
073002 (2005)].
In addition and counter-intuitively, it is shown that for large initial 
vibrational excitation ionization takes place favorably at large 
angles between the laser polarization and molecular axis. A local 
ionization model delivers a transparent explanation of these findings.
\end{abstract}

\maketitle

\newpage

Even for  nature's simplest molecule $\text{H}_2^+$, no full-dimensional 
description (including rotation of the nuclei) of the ionization dynamics 
has been given up to date. While in the early days important mechanisms 
of strong field ionization could be identified in calculations with frozen 
nuclei \cite{gimi+95}, it was shown subsequently that nuclear vibration  
does have a crucial influence on the ionization dynamics. Prominent effects 
include the washout of charge resonance enhanced ionization  at specific 
internuclear distances (CREI) \cite{Bandrauk95,PhysRevLett.79.2022} by  
vibrational motion \cite{PhysRevA.57.1176,PhysRevA.78.063419} as well as 
Lochfra\ss \ (internuclear separation dependent ionization) 
\cite{PhysRevLett.97.103003,PhysRevLett.97.103004}. For laser pulses much 
shorter than a typical rotational period it has been reasonably assumed 
that rotation can be neglected \cite{PhysRevA.52.2977,1367-2630-11-8-083014}.
However, recent theoretical work \cite{0953-4075-42-9-091001} has revealed 
the importance of rotation for the dissociation channel even for very short 
laser pulses. From a key experiment \cite{PhysRevLett.95.073002} it could be 
concluded that rotation and/or orientation is also relevant for ionization, 
 which appears to be overestimated when the molecule remains aligned with the 
laser polarization. 
 
 All these findings can be put into a stringent perspective with the results 
from our present calculations taking all degrees of freedom into account.
The mixed quantum-classical non-adiabatic quantum molecular dynamics (NA-QMD) 
method \cite{1Kunert03,Uhlmann06,tobepub} which has enabled us to perform a 
full dimensional calculation is still approximate since it treats the nuclei 
classically. Hence, we have carefully assessed its validity for parameter 
regimes where accurate quantum calculations can be performed. In  
\cite{1367-2630-11-8-083014} dissociation and ionization of laser-aligned 
H$_2^+$ molecules were calculated  and in \cite{tobepub} we have obtained 
vibrationally resolved angular distributions of dissociated fragments for 
laser parameters where ionization can be neglected. In both cases we have found 
quantitative agreement with appropriate quantum results, in the first case from 
a dimensionally reduced quantum approach \cite{1367-2630-11-8-083014}, in the 
second case from full-dimensional quantum calculations, however restricted to 
dissociation \cite{tobepub}. Therefore, the mixed quantum-classical NA-QMD 
approach appears to be well suited to quantify ionization for $\text{H}_2^+$ 
including all degrees of freedom, for which no calculations exist so far.

Our starting point is the time-dependent electronic Schr\"odinger equation 
(atomic units are used)
\begin{align}
\text{i} \frac{\partial}{\partial t} 
&\Phi(\mathbf{r},t;\{\mathbf{R}_A\})	
=
\nonumber
\\
& \left[\hat{T}_{\mathbf{r}}\!- \! \sum\limits_{A=1}^2\frac{1}{\mid \! \mathbf{r} 
\! - \! {\mathbf{R}}_A \! \mid} \! + \! z \epsilon(t)\right]  
\Phi(\mathbf{r},t;\{\mathbf{R}_A\}) \text{,}
\label{TDSEel}
\end{align} 
which is solved in basis expansion self-consistently with Newton's equations of 
motion for the nuclei.  Here, $\mathbf{R}_A$ ($A=1,2$) and $\mathbf{r}$ denote 
the nuclear and electronic coordinates, $\hat T_\mathbf{r} = -1/2\nabla^{2}_{\mathbf r}$  
the electronic kinetic energy, and $\epsilon(t)$ is the electric field of the laser.
Details regarding the method and its numerical implementation can be found in 
\cite{tobepub,Uhlmann06}. At the final time $t_\text{f}$ the vibrationally resolved 
ionization (I) and dissociation (D) probabilities are constructed from probability 
densities along relevant nuclear trajectories,
\begin{align}
 P_{\rm I}(\nu)&= 1-Z_\nu^{-1} \sum_{i=1}^{Z_\nu}  N[\mathbf R_i](t_\text{f}) \\
 P_{\rm D}(\nu)&= Z_\nu^{-1}  \sum_{i: R_i(t_\text{f}) > R_{\rm D}}^{Z_\nu}  
N[\mathbf R_i](t_\text{f}) \text{,}
\end{align}
where $N[\mathbf R](t)$ is the norm of the electronic wavefunction, which is a functional 
of the trajectory $\mathbf R(t)$ and a function of time. Its decrease in time, due to the 
presence of an absorber potential \cite{tobepub,Uhlmann06}, is a measure of ionization. 
Trajectories with final internuclear distance $R>R_\text{D}=10 \text{\ a.u.}$ contribute 
to dissociation. $Z_\nu$ denotes the number of trajectories, sampled according to a 
micro-canonical distribution of a given initial vibrational state $\nu$. With  the molecule 
initially in its  rotational ground state, the nuclear rotation angle $\theta$ is sampled 
uniformly in the interval $[0,\frac{\pi}{2}]$, where one has to take into account the 
additional weighting factor $\sin{\theta_i}(t_0)$ \cite{tobepub}. Complementary to the 
full dimensional calculations we have also performed dynamical calculations but with 
fixed nuclear orientation (frozen rotation) in order to study the influence of the 
rotational dynamics on the results.

We have determined vibrationally resolved ionization and  dissociation probabilities  
for a linearly polarized laser pulse with an amplitude 
$\epsilon(t) = \epsilon_{0}\sin^{2}(\pi t/T)\cos\omega t$ at a wavelength of 
$800 \text{\ nm}$ (corresponding to $\omega \approx 0.057 \text{\ a.u.}$), peak 
intensity of $I = 2 \cdot 10^{14} \frac{\text{W}}{\text{cm}^2}$ (corresponding to 
$\epsilon_{0} \approx 0.075 \text{\ a.u.}$) and total pulse length of $T = 50 \text{\ fs}$. 
Similar laser parameters were used in recent experiments 
\cite{PhysRevLett.85.4876,PhysRevLett.95.073002,PhysRevA.74.043411}.

\begin{figure}[htb]
\begin{center}
	\parbox{0.45\textwidth}{\includegraphics[width=0.45\textwidth]{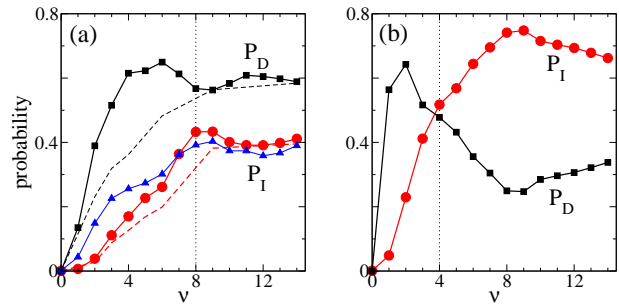}}
\end{center}
\caption{(Color online) Ionization probability $P_{\rm I}(\nu)$ (red circles), and dissociation 
probability $P_{\rm D}(\nu)$ (black squares) for (a) initially rotationally cold molecules 
(isotropically orientated) and (b) \emph{laser-aligned}  molecules (see also 
\cite{1367-2630-11-8-083014}) as a function of the initial vibrational state $\nu$. The dashed lines 
in (a) represent results for  fixed nuclear orientation (frozen rotation). The vertical line denotes 
the vibrational level, above which fragmentation saturates, $P_\text{D}+P_\text{I}=1$. Finally, the 
ionization probability based on the LIM (Eq.\ \ref{equ:rate})  is shown with blue triangles.}
\label{fig:diss+ion_moving}
\end{figure}

Fig.~\ref{fig:diss+ion_moving}a reveals that the full dimensional integral results for 
ionization (red circles, including rotational dynamics) agree quite well with those for 
frozen nuclear axis (dashed lines), where in both calculations the molecules are 
isotropically distributed in the beginning. This is understandable since over the length 
of the laser pulse (50 fs) during which ionization is possible the molecule hardly rotates. 
Dissociation, however, is much slower and its probability consequently deviates considerably 
from the result for frozen angles (black squares versus black dashed line). Only for weak 
binding at large $\nu$ dissociation happens even more quickly than ionization so that 
$P_\text{D}$  agrees with the frozen result. Moreover, this implies that beyond fragmentation 
saturation ($P_\text{I} + P_\text{D} = 1$, vertical dotted line),  ionization is suppressed 
by dissociation according to $P_\text{I}(P_\text{D})=1-P_\text{D}$. Finally, the two maxima 
in $P_\text{D}$ around $\nu=4$ and $\nu=11$ above the frozen probability are due to the well 
known dressed two- and one-photon states, respectively  \cite{tobepub}.

\begin{figure}[t]
\begin{center}	
	\includegraphics[width=0.3\textwidth]{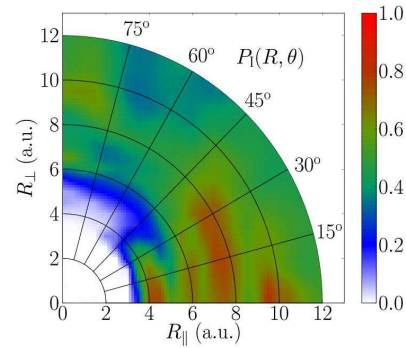}
\end{center}
\caption{(Color online) The ionization probability $P_{\rm I}(\mathbf R)=1-N(\mathbf R,T)$ calculated 
for fixed nuclear positions with the parallel and perpendicular component $R_\parallel = R  \cos{\theta}$ 
resp. $R_\perp= R \sin{\theta}$. Geometric alignment roughly manifests itself in increasing ionization 
probabilities towards the laser polarization direction. Near laser polarization direction clearly a 
region of strong ionization is visible, in particular the $1$-photon-resonance at 
$R \approx 4.3 \ \text{a.u.}$ as well as additional enhanced ionization maxima related to CREI.}
\label{fig:ion_fixed}
\end{figure}

The corresponding probabilities for the molecule aligned with the laser polarization 
(Fig.~\ref{fig:diss+ion_moving}b) differ quantitatively and qualitatively from the full 
dimensional analogs (Fig.~\ref{fig:diss+ion_moving}a). Due to the much stronger ionization, 
fragmentation saturation sets in already near $\nu = 4$. Hence, for the aligned molecule 
ionization suppresses dissociation for high $\nu$, according to 
$P_\text{D}(P_\text{I})= 1 - P_\text{I}$, in striking contrast to the real case 
(Fig.~\ref{fig:diss+ion_moving}a). For the highest vibrational levels ionization decreases 
again, because there is a large probability that the molecules  have an internuclear distance 
outside the strong ionization region (Fig.~\ref{fig:ion_fixed}) when the laser reaches peak 
intensity. From Fig.~\ref{fig:ion_fixed} it is also clear why the full dimensional and laser 
aligned molecular response to the light pulse is so different: The coupling to the light 
changes considerably for different alignment angles and it is strongest in the aligned case 
where molecular enhancement mechanisms such as dressed state resonances and CREI are operative.  

\begin{figure}[h]
\begin{center}
 \parbox{0.55\textwidth}{ \includegraphics[width=0.55\textwidth]{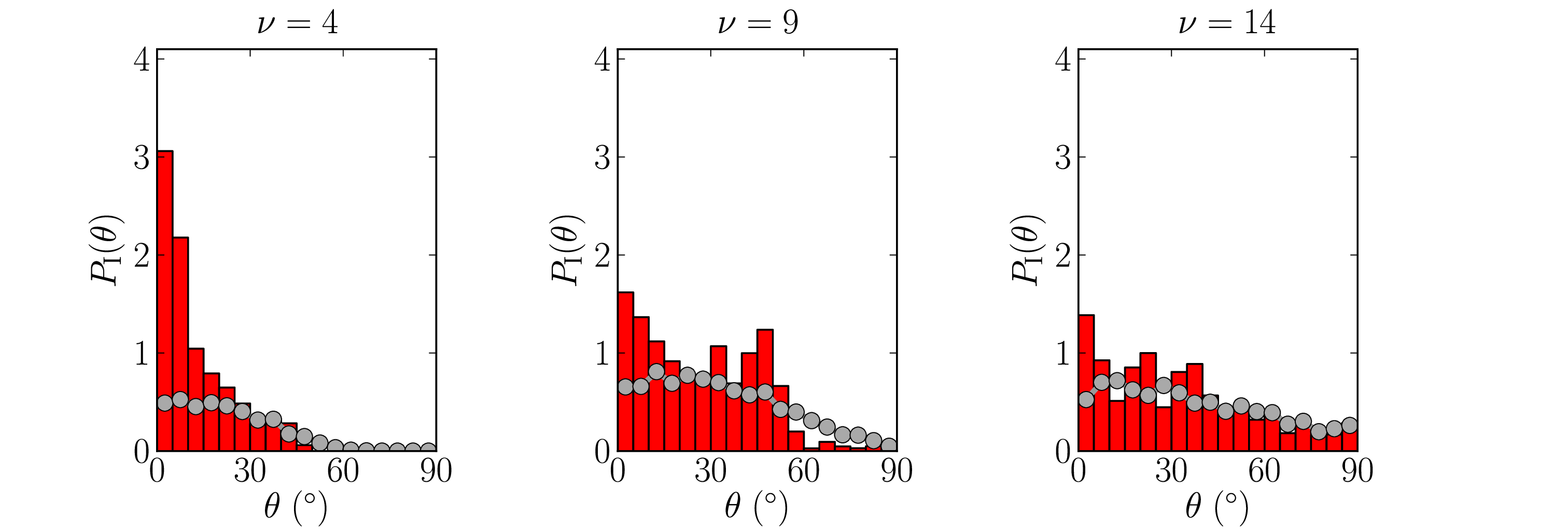}} 
 \parbox{0.55\textwidth}{ \includegraphics[width=0.55\textwidth]{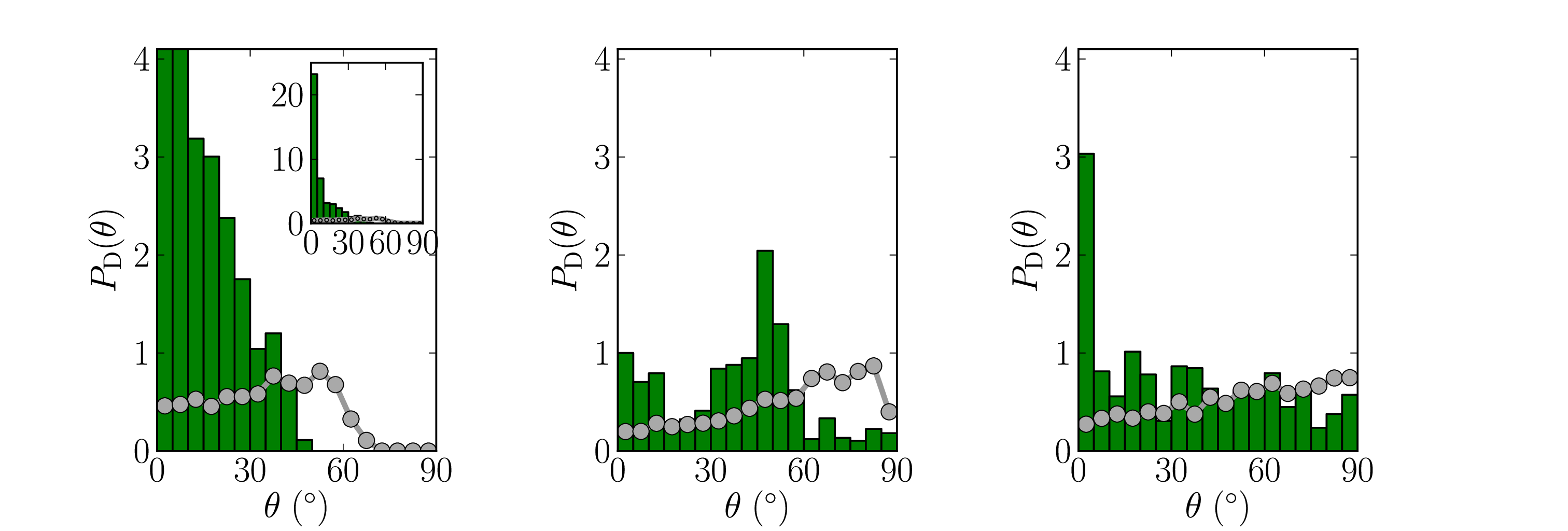}}
\end{center}
\caption{(Color online) Angular distribution of   fragments   starting from vibrational H$_2^+$ 
levels $\nu = 4,\ 9,\ {\rm and}\ 14$ for ionization $P_{\rm I}(\theta)$ (red bars) and dissociation 
$P_{\rm D}(\theta)$ (green bars) from  full dimensional calculations (cf.~\cite{tobepub} for 
definition of $P_{\rm D}(\theta)$, $P_{\rm I}(\theta)$ analogous). The results calculated with 
frozen rotation  are given with gray circles.}
\label{fig:ion_theta}
\end{figure}
The angular distributions of dissociated and ionized fragments  (Fig.~\ref{fig:ion_theta}) confirm 
what we have concluded from the integral probabilities of Fig.~\ref{fig:diss+ion_moving}. For the 
ionization channel, the angular distributions follow roughly the behavior  expected from geometric 
alignment. In contrast, the angular distributions for dissociation from full dimensional and 
rotationally frozen calculations exhibit  an opposite trend for increasing angle. Moreover,
the strong alignment (for low and high  $\nu$) and anti-alignment  for intermediate $\nu$ which 
has been explained and understood within the Floquet picture (e.g., \cite{tobepub}) is missing 
for frozen angles.

\begin{figure}[t]
\begin{center}
 \includegraphics[width=0.5\textwidth]{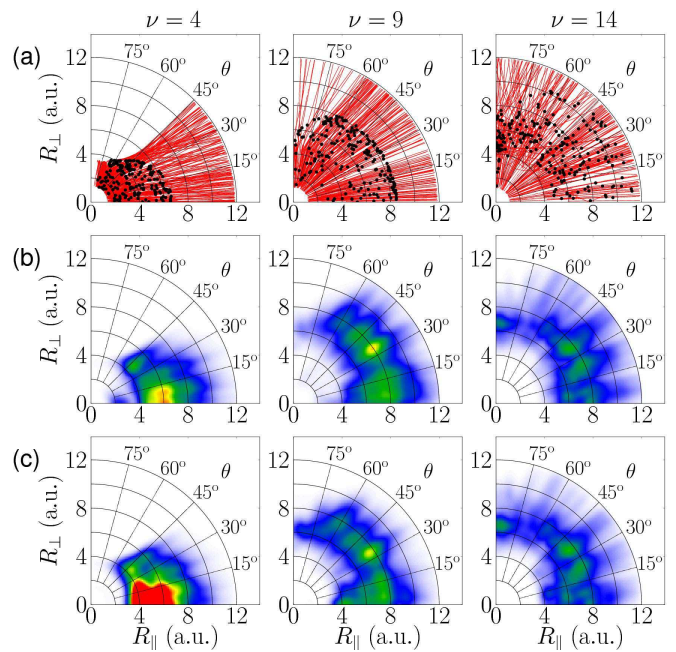}
\end{center}
\caption{(Color online) (a) Trajectories of fragmenting molecules (red lines) as well as nuclear 
positions at peak intensity (black dots) for exemplary initial vibrational levels ($\nu = 4,9,14$). 
(b) Corresponding exact ionized density $\rho_{\rm I}(\mathbf R)$ and (c) ionized density in the 
LIM $\rho_{\rm I}^{\rm LIM}(\mathbf{R})$ (linear color scale). The nuclear coordinates are the same 
as in Fig.~\ref{fig:ion_fixed}.}
\label{fig:comparison}
\end{figure}

Summarizing the interpretation of the results so far it has become clear that a full dimensional 
approach is necessary to capture the dynamics adequately. Nevertheless, as the ionization probability 
depends exponentially on the laser intensity, a simplified dynamical description may be possible. In 
our case $P_\text{I}$ from Fig.~\ref{fig:ion_fixed} is reduced ten times when using half the peak 
intensity which is also observed experimentally  \cite{PhysRevLett.95.073002}. Hence, the full 
dynamical ionization should  mainly be determined by the nuclear positions around peak intensity 
(25 fs into the pulse), dependent on the initial vibrational state as illustrated in 
Fig.~\ref{fig:comparison}a. To understand  the influence of nuclear motion on ionization, we define 
the ionized nuclear density 
\begin{align}
 \rho_{\rm I}(\mathbf{R}) = -Z_\nu^{-1} \sum_{i=1}^{Z_\nu}\int\limits_{0}^{t_\text{f}} \text{d}t\   
\dot{N}[\mathbf R_i](t)\ \delta(\mathbf{R}-\mathbf R_{i}(t)) \text{\ ,}
\label{equ:exact}
\end{align}
which reveals preferential nuclear positions for ionization (Fig.~\ref{fig:comparison}b).
Surprisingly, for large $\nu$ ionization proceeds favorably at large angles.
It also corroborates the ``naive'' picture of Fig.~\ref{fig:comparison}a that ionization mainly takes 
place at the time of peak intensity and is therefore sensitive to the dynamically reached nuclear 
positions at this time.
 
To quantify this observation, we may assume that ionization happens instantaneously neglecting memory 
effects for nuclear motion in the approximation
\begin{align}
 \dot N[\mathbf{R}_i](t) \approx -\Gamma(\mathbf{R}_i(t),t)\ N[\mathbf{R}_i](t)
 \label{LIR}
\end{align}
with an instantaneous local ionization "rate" $\Gamma(\mathbf R,t)=-{\dot N(\mathbf R,t)}{/N(\mathbf R,t)}$.
Note that this assumption is not trivial, since the ionization of any trajectory in the 
swarm depends in principle on its whole time evolution.
The instantaneous local ionization rate $\Gamma(\mathbf R,t)$ 
is calculated within the fixed-nuclei approximation using the same laser conditions as in the full 
dynamical calculations. This rate is not a simple function of time, internuclear distance and alignment 
angle, but also depends on the pulse shape of the laser. Inserting \eqref{LIR} into \eqref{equ:exact} 
leads to a  local ionization model (LIM), where the ionized nuclear density takes the product form
\begin{align}
\rho_{\rm I}^{\rm LIM}(\mathbf{R}) = 
\int\limits_{0}^{t_\text{f}} \text{d}t\ \Gamma(\mathbf R,t)\ \rho(\mathbf{R},t) \text{}
\label{equ:rate}
\end{align}
with the time-dependent nuclear density
\begin{align}
\rho(\mathbf{R},t)= Z_\nu^{-1} \sum_{i=1}^{Z_\nu}  N[\mathbf R_i](t)\ \delta(\mathbf{R}-\mathbf R_{i}(t)) 
\text{\ .}
\end{align}
As one can see from Fig.~\ref{fig:comparison}c, the separation
of  nuclear dynamics and ionization according to \eqref{equ:rate} reproduces the full result
Fig.~\ref{fig:comparison}b quite well and is in accord with the spirit of  \cite{PhysRevLett.103.183601}
with the difference that our ionization rates are extracted from the exact laser pulse.

Note, however, that  the LIM overestimates ionization for low $\nu$, which is due to the proximity 
of a 1-photon resonance close to $R \approx 4.3 \ \text{a.u}$. For higher $\nu$ the model reproduces 
that ionization takes place dominantly at intermediate angles as a consequence of dissociation dynamics 
\cite{tobepub}. The quantitative accuracy of the LIM (Fig.~\ref{fig:diss+ion_moving}a, blue triangles) 
is overall comparable to the rotationally frozen result  (dashed line), however LIM better reflects 
dynamical details, such as the influence of dressed state resonances.
\begin{figure}[htb]
\begin{center} 
\includegraphics[width=0.5\textwidth,height=4cm]{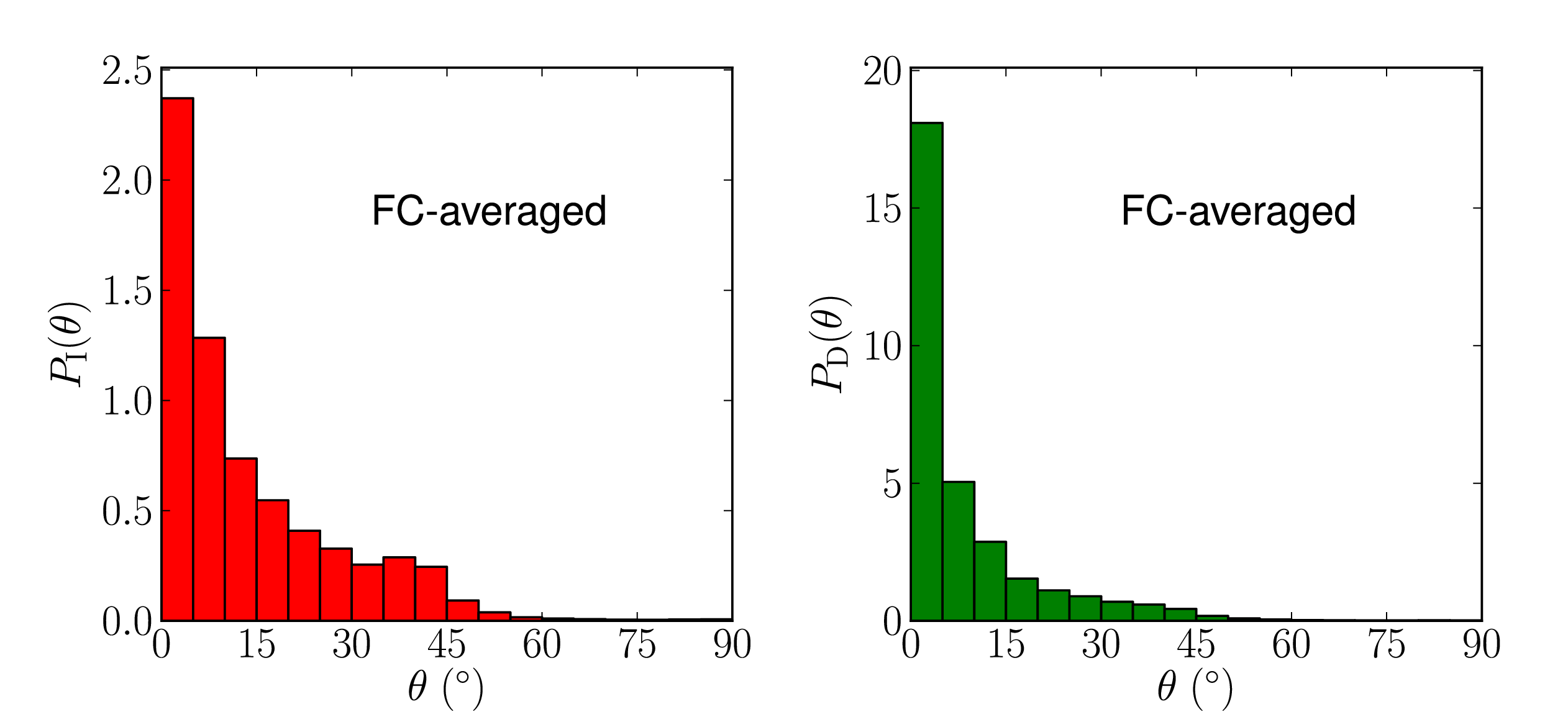}
\end{center}
\caption{(Color online) Franck-Condon averaged angular distribution for ionization ($P_\text{I}(\theta)$) 
and dissociation ($P_\text{D}(\theta)$).}
\label{fig:fcaverage}
\end{figure}

A serious 
quantitative comparison with the experiment \cite{PhysRevLett.95.073002}, which reveals one order of 
magnitude difference between total  ionization and dissociation yield, requires Franck Condon averaging 
over initial vibrational levels and in addition focal volume and thermal averaging. While the former 
is a generally valid result, the latter depends on the exact conditions of individual experiments, 
which differ and are not trivial to quantify. Hence, we restrict ourselves to calculate full Franck-Condon 
averaged \cite{FCF} angular distributions for ionization and dissociation and discuss modifications 
due to volume averaging qualitatively. As one can see in Fig.~\ref{fig:fcaverage}, $P_\text{D}(\theta)$ 
is more strongly aligned along the laser polarization than $P_\text{I}(\theta)$ in seeming discrepancy 
with the opposite observation in the experiment. However, as noted in \cite{PhysRevLett.95.073002}, 
the dynamics changes substantially with small changes in pulse length  if the latter is close to the 
vibrational time scale which is the case here. Hence,  the small differences in  experimental and 
theoretical pulse length and shape may matter. Secondly, the spatial intensity profile of the laser 
will favor contributions of ionized fragments in the wings of the laser focus which originate 
dominantly from laser-aligned molecules.  In the experiment \cite{PhysRevLett.95.073002} low laser 
intensities contribute considerably to the signal due to the width of the ion beam used. On the other 
hand,  $P_\text{D}(\theta)$ would broaden for large contributions from lower intensities (cf. the IDS 
procedure used in  \cite{Wang:05}). The total  dissociation and ionization probabilities from our 
calculation, $P_\text{I} = 0.13$ and $P_\text{D} = 0.41$,  reproduce the experimental trend, that 
ionization is considerably smaller than dissociation in sharp contrast to predictions from dimensionally 
reduced calculations.

In summary, we have presented a complete study of strong field ionization and dissociation of H$_2^+$. 
All nuclear and electronic degrees of freedom have been included in the framework of the mixed 
quantum-classical NA-QMD method whose applicability has been checked carefully in prior work by  
comparison to quantum results where available. It is worthwhile to note, that NA-QMD \cite{1Kunert03,Uhlmann06} 
is also applicable to polyatomic many-electron systems \cite{PhysRevLett.98.058302}.

We have found, that nuclear rotation enhances dissociation so that  $P_\text{D}>P_\text{I}$ for 
$\text{H}_{2}^{+}$ under short pulses  in agreement with the experiment \cite{PhysRevLett.95.073002} 
but in discrepancy with dimensionally reduced calculations. Furthermore, vibrationally resolved 
ionization $P_\text{I}(\nu)$  is reasonably well reproduced with frozen rotation, while 
$P_\text{D}(\nu)$ is enhanced in comparison to frozen nuclear geometry with two maxima whose origin 
can be easily traced to one- and two-photon dressed states. The latter connection is almost completely 
lost in the dimensionally reduced result since due to the strong ionization, fragmentation saturation 
occurs already for $\nu > 4$ and leads to suppression of dissociation.

We conclude that full dimensional calculations are necessary for short pulses to obtain even 
qualitatively correct results. Nevertheless, a simplification arises  in a local ionization model 
(LIM) from the separation of ionization and nuclear dynamics, taking advantage of the fact that 
ionization is fast, but happens dominantly at peak intensity,  when nuclei have already moved. 
Finally, we could demonstrate that rotationally frozen dynamics  is a good approximation when 
dissociation or ionization happens fast with respect to the rotational time scale. This is always 
the case for ionization with a 50 fs pulse and applies regarding dissociation to the highest vibrational 
levels, which dissociate before the molecule rotates substantially.

We gratefully acknowledge the allocation of computer resources from the ZIH of the TU Dresden and 
support by the DFG through Grants No. SCHM-957 and SCHU-1557.


\end{document}